\documentclass[prl,twocolumn,showpacs,preprintnumbers,amsmath,amssymb]{revtex4}

\usepackage{graphicx}
\usepackage{amsmath}
\usepackage{amsbsy}
\usepackage{dcolumn}
\usepackage{bm}

\begin{document}

\title{$0-\pi$ transitions in Josephson junctions with antiferromagnetic interlayers}

\author{Brian M. Andersen,$^1$ I. V. Bobkova,$^2$ P. J. Hirschfeld,$^1$ and Yu. S. Barash$^2$}
\affiliation{$^1$Department of Physics, University of Florida,
Gainesville, Florida 32611-8440, USA\\ $^2$Institute of Solid
State Physics, Russian Academy of Sciences, Chernogolovka, 142432 Russia}

\date{\today}

\begin{abstract}
We show that the dc Josephson current through
superconductor-antiferromagnet-superconductor (S/AF/S) junctions
manifests a remarkable atomic scale dependence on the interlayer
thickness. At low temperatures the junction is either a $0$- or
$\pi$-junction depending on whether the AF interlayer consists of
an even or odd number of atomic layers. This is associated with
different symmetries of the AF interlayers in the two cases. In
the junction with odd AF interlayers an additional $\pi-0$
transition can take place as a function of temperature. This
originates from the interplay of spin-split Andreev bound states.
Experimental implications of these theoretical findings are
discussed.
\end{abstract}


\maketitle The developing field of superconducting spintronics
subsumes many fascinating physical phenomena with potential
applications that may complement non-superconducting spintronic
devices\cite{sarma04}. In addition there is an increasing interest
in the novel properties of interfaces and junctions of
superconductors and magnetic materials\cite{efetov05,buzdin05}. An
important example is the $0-\pi$ transition in
superconductor-ferromagnet-superconductor (S/F/S) junctions where,
depending on the temperature and the width of the ferromagnetic
interlayer, the ground state of the junction is characterized by
an intrinsic phase difference of $\pi$ between the two
superconductors \cite{buzdin05,ryazanov01,aprili02}. This may be
important for superconducting digital circuits, and has been
proposed as a possible basis for quantum
qubits\cite{beasley,blatter,zagoskin,ustinov,yamashita}. The
interplay of magnetic and superconducting order leads to
interesting mesoscopic physical phenomena also at interfaces
between antiferromagnets and superconductors. Recent theoretical
studies have shown that a characteristic spin dependent
quasiparticle reflection at the AF surface, the so-called
Q-reflection, combined with Andreev reflection on the
superconducting side, leads to novel low-energy bound states near
such interfaces\cite{bobkova05}. These Andreev bound states have
important consequences for the associated proximity effect and
can, for example, be detected in tunneling spectroscopy as sub-gap
peaks in the resulting local density of states\cite{andersen05}.

In this Letter we investigate effects of the Q-reflection on the
Josephson current in S/AF/S tunnel junctions in $s$-wave
superconductors. An enhancement of the low-temperature critical
current in such junctions was found in the limit of a sufficiently
small ratio $m/t$, where $t$ and $m$ denote the hopping matrix
element and the antiferromagnetic order parameter,
respectively\cite{bobkova05}. As shown below, the parameter $m/t$
becomes sufficiently small in tunnel junctions only under the
condition $(m/t)\ll \sqrt{D}$, where $D$ is the transparency
coefficient. We will find that in the opposite case $(m/t)\gtrsim
\sqrt{D}$, which is also realistic for fabrication, the Josephson
current in S/AF/S junctions exhibits new interesting properties.
At low temperature $T$, the current-phase relation reveals either
a $0$- or a $\pi$-junction state depending on whether the AF
interlayer consists of an even or odd number of atomic layers,
respectively. This atomic-scale thickness dependence differs from
that of S/F/S junctions where the period of the alternation
between $0$- and $\pi$-states strongly depends on the exchange
field which controls the proximity-induced damped spatial
oscillations of the pairing amplitude in the ferromagnetic metal.
In contrast, for the S/AF/S junctions the $0-\pi$ behavior is a
true even/odd effect related to the difference in symmetries of
the odd and even AF interfaces and the corresponding interface
$\cal S$ matrices, as well as the spectra of the Andreev bound
states. For odd (110) AF interlayers, the supercurrent displays a
prominent anomaly with increasing $T$ {\sl revealing another
$\pi-0$ transition}. We will show that this remarkable result is a
consequence of the interplay of spin-split Andreev bound states
contributing with opposite sign to the total supercurrent. Lastly,
we discuss possible experimental consequences of the effects in
question. Our theoretical analysis includes both the
self-consistent numerical solutions of the Bogoliubov-de Gennes
(BdG) equations and a quasiclassical analytical approach to the
superconducting leads, allowing us to describe the AF interface
fully microscopically with atomic scale accuracy, and interpret
the results physically.

{\it Model}. The Hamiltonian is defined on a 2D square lattice
with superconducting $\Delta_{i}$ and magnetic $m_i$ order
parameters, and lattice constant $a=1$:
\begin{eqnarray}\label{hamiltonian}
\hat{H}= &-& t \sum_{\langle ij \rangle\sigma}
\hat{c}_{i\sigma}^{\dagger}\hat{c}_{j\sigma} + \sum_{i} \left(
\Delta_{i}
\hat{c}_{i\uparrow}^{\dagger}\hat{c}_{i\downarrow}^{\dagger} +
\mbox{H.c.} \right) \nonumber\\ &-& \sum_{i\sigma} \mu
\hat{n}_{i\sigma} + \sum_{i} m_i \left(\hat{n}_{i\uparrow} -
\hat{n}_{i\downarrow} \right).
\end{eqnarray}
Here, $\hat{c}_{i\sigma}^{\dagger}$ creates an electron of spin
$\sigma$ on the site $i$, $\mu$ is the chemical potential and
$\hat{n}_{i\sigma}=\hat{c}_{i\sigma}^{\dagger}\hat{c}_{i\sigma}$
is the particle number operator.
The associated BdG equations are
\begin{equation}\label{BdG}
\sum_j \left( \begin{array}{cc} {\mathcal{K}}^{+}_{ij,\sigma}&
{\mathcal{D}}_{ij,\sigma} \\
{\mathcal{D}}^*_{ij,\sigma} & -{\mathcal{K}}^{-}_{ij,\sigma}
\end{array}\!\right)\! \left( \begin{array}{c} u_{n\sigma}(j) \\
v_{n\overline{\sigma}}(j) \end{array}\!\right)\! =\!
E_{n\sigma}\! \left( \begin{array}{c} u_{n\sigma}(i) \\
v_{n\overline\sigma}(i) \end{array}\! \right).
\end{equation}
The diagonal blocks are given by
${\mathcal{K}}^{\pm}_{ij}=-t\delta_{\langle
ij\rangle}-\mu\delta_{ij}\pm\sigma m_i\delta_{ij}$, where
$\sigma=+1/-1$ for up/down spin and $\delta_{ij}$ and
$\delta_{\langle ij \rangle}$ are the Kronecker delta symbols
connecting on-site and nearest neighbor sites, respectively. The
off-diagonal block ${\mathcal{D}}_{ij}$ describes $s$-wave pairing
${\mathcal{D}}_{ij}=-\Delta_i\delta_{ij}$. The net magnetization
$M_i=\frac{1}{2} \left[ \langle \hat{n}_{i\uparrow} \rangle\ -
\langle \hat{n}_{i\downarrow} \rangle \right]$ and the pairing
amplitude $F_{i}=\langle\hat{c}_{i\downarrow}
\hat{c}_{i\uparrow}\rangle$ are related to $m_i$ and $\Delta_{i}$
by $m_i=U_i M_i$ and $\Delta_{i}=-V_i F_{i}$. The coupling
constants $U_i$ ($V_i$) are site dependent and non-zero on (off)
the $L$ atomic chains constituting the AF interlayer. This
stabilizes the staggered AF order on the interlayer and the
superconducting order outside this region. We choose the $x$ ($y$)
axis to run perpendicular (parallel) to the interface. The
calculations for planar tunnel junctions with crystal periodicity
along the interface are reduced to a 1D problem by Fourier
transforming along $y$. This introduces a crystal-vector component
$k_y$ as a parameter and the BdG equations have to be diagonalized
for each $k_y$.

The dc Josephson current $j_{rr'}$ between two neighboring sites
$r=(x,y)$ and $r'=(x',y')$ is $j_{rr'}=-(iet/\hbar) \sum_\sigma
\left[ \langle \hat{c}_{r\sigma}^{\dagger}\hat{c}_{r'\sigma}
\rangle - \langle \hat{c}_{r'\sigma}^{\dagger}\hat{c}_{r\sigma}
\rangle \right]$. Below we report the results for the current per
unit length $j_{xx'}=(1/l_y)\sum_y j_{rr'}$ obtained by summing
along the interface of length $l_y$ over all neighboring links
between $x$ and $x'$ chains near the interface. For the (110)
interface $x'=x+(1/\sqrt{2})$ and we get
\begin{eqnarray}\label{currentfinal}\nonumber
j_{xx'}&=& -\frac{2iet}{\hbar l_y} \sum_{k_yn\sigma}
\cos\left(\frac{k_y}{\sqrt{2}}\right) \left[
u_{n\sigma}^*(x)u_{n\sigma}(x')f(E_{n\sigma}) \right. \nonumber\\
&+& \left.
v_{n\overline\sigma}(x)v^*_{n\overline\sigma}(x')f(-E_{n\sigma}) -
(x\leftrightarrow x') \right],
\end{eqnarray}
where $-{\pi}/{\sqrt{2}} < k_y \leq {\pi}/{\sqrt{2}}$ and $f(E)$
is the Fermi function. For the (100) junction $x'=x+1$ and the
current takes the form (\ref{currentfinal}), but without the
$\cos(k_y/\sqrt{2})$ factor and with $-\pi < k_y \leq\pi$. Since
the vector potential generated by the current is not taken into
account, we discuss only planar junctions in the tunneling limit
when the current $j$ is well below the thermodynamic critical
pair-breaking current. Below, we fix the phase of the
superconducting order parameter at each end of the system, obtain
the self-consistent solutions, and calculate the current based on
Eq.(\ref{currentfinal}). As is well-known, the current is only
conserved when the superconducting order parameter is calculated
fully self-consistently\cite{BTK}. For more details on the
numerical and analytical approaches used in this Letter, we refer
the reader to Ref. \onlinecite{andersen05}.

\begin{figure}[!tbh]
\includegraphics[width=7.8cm]{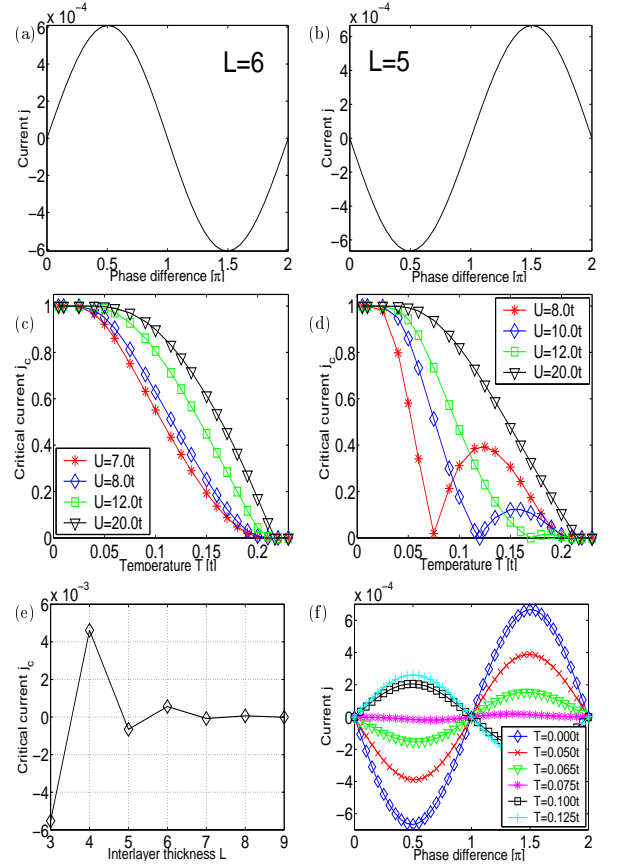}
\caption{(Color online) Current-phase relation at $T=0$, $U=8.0t$
(a,b) and temperature dependence of the critical current
$j_c=|j(\frac{\pi}{2})|$ (c,d) for S/AF/S (110) junctions with
$L=6$ (a,c) and $L=5$ (b,d). In (c,d) the graphs have been
normalized to their values at $T=0$. (e) Oscillations of the
critical current as a function of the AF thickness $L$ ($U=8.0t$).
(f) Current-phase relation at various temperatures along the
$U=8.0t$ curve in (d). Parameters used in all figures: $V=2.0t$,
$\mu=0$. Currents are in units of $et/\hbar
l_y$.\label{currentphaseT0vsafx}}
\end{figure}

{\it Results}. We have studied the dc Josephson current in both
(100) and (110) S/AF/S junctions. Below, however, we focus on the
(110) interfaces where the effects in question are more
pronounced. Spins along (across) a (110) AF interlayer are
identically aligned (alternate). Fig. \ref{currentphaseT0vsafx}a,b
show two representative $T=0$ current-phase relations for the
S/AF/S tunnel junctions with different thicknesses $L$ of the
sandwiched (110) AF layer. The curves display striking $0$- or
$\pi$-junction sinusoidal behavior depending on whether the AF
interface contains an even (Fig. \ref{currentphaseT0vsafx}a) or
odd (Fig. \ref{currentphaseT0vsafx}b) number of chains,
respectively. Fig. \ref{currentphaseT0vsafx}e shows the critical
current as a function of thickness $L$. The temperature dependence
of the critical current $j_c(T)$ is shown in Fig.
\ref{currentphaseT0vsafx}c,d for varying magnetic strength $U$.
Clearly, the magnetism leads to remarkable anomalous $T$
dependence of the critical current. For small values of $U$, where
the influence of the Q-reflection is more pronounced since the
Andreev bound states have lower energy, $j_c(T)$ deviates from the
standard behavior by varying monotonically for the even junctions
and non-monotonically for odd $L$. In particular, we find that in
addition to the low $T$ alternating $0-\pi$ transitions as a
function of $L$, the odd junctions can exhibit another $\pi-0$
transition as a function of $T$. This $\pi-0$ transition is
clearly shown in Fig. \ref{currentphaseT0vsafx}f, where we plot
the current-phase relation at various $T$ along the $U=8.0t$ curve
in Fig. \ref{currentphaseT0vsafx}d. Note that in Fig.
\ref{currentphaseT0vsafx}c,d the conventional behavior near $T_c$,
$j_c \sim [ 1-T/T_c ]$, is recovered only in the limit of large
$U$. The anomalous temperature dependence near $T_c$ appears to be
associated with the magnetism leaking into the superconducting
leads by the proximity effect\cite{andersen05}.

\begin{figure}[t]
\includegraphics[width=8.5cm,height=5.0cm]{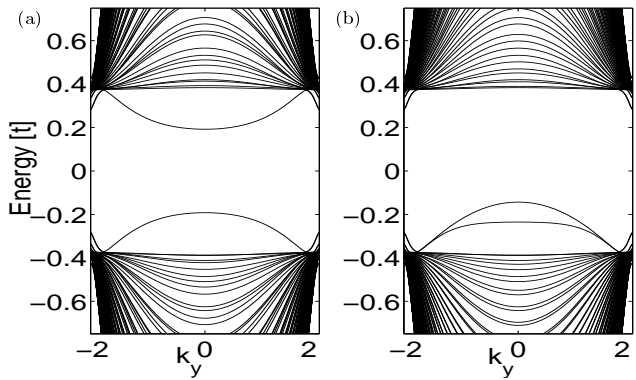}
\caption{Eigenbands for the S/AF/S junctions with $U=8.0t$ and
$\mu=T=\varphi=0.0$ and $L=6$ (a) and $L=5$ (b). As seen, the even
or odd number of magnetic layers lead to qualitatively different
bound state bands inside the gap.\label{bandskyfig}}
\end{figure}

In junctions between identical superconductors with a thin
interlayer $L\ll\xi_s\sim\hbar v_F/|\Delta|$, the whole Josephson
current is carried through the interface by the phase-dependent
subgap Andreev bound states. The bound states at the two separate
S/AF interfaces\cite{bobkova05,andersen05}, mix in the junction
geometry resulting in qualitatively different bound state bands
for the odd or even interlayers. This is seen in Fig.
\ref{bandskyfig} where we show the spin-down eigen-spectrum for a
even (a) and odd (b) (110) S/AF/S junction as a function of $k_y$.
In the case of odd $L$ only the total spectrum, including both
spin-down and spin-up states, is symmetric with respect to the
Fermi surface. For larger values of $U$, the transparency of the
junction decreases, the specular reflection becomes more
pronounced and the bound states move towards the gap edge and
become more extended.

We have carried out analytical calculations of the subgap spectrum
under the conditions $\Delta\!\ll\! m,t$,\, $L\!\ll\xi_s$, whereas
our numerical self-consistent studies are applicable in a wider
regime. The normal-state spin-dependent reflection and
transmission amplitudes are the same on both sides of a (110) odd
AF interface: $r_{\sigma}=\sqrt{ R_{odd}}\exp(i\sigma\Theta/ 2)$,
$d_{\sigma}=i\sigma\sqrt{D_{odd}}\exp\left(i\sigma\Theta/2\right)$.
Here the transparency coefficient is spin-independent $D_{odd}(
k_y)=\cosh^{-2}\left[2(L+\frac{1}{2})\beta\right]$,\, $\sinh\beta
=m/[4t\cos(k_y/\sqrt{2})]$ and $R_{odd}(k_y)=1-D_{odd}(k_y)$. The
difference $\Theta=\Theta_{\uparrow}-\Theta_{\downarrow}$ between
phases of spin-up
and spin-down reflection amplitudes takes the form $\sin\Theta(k_y
)=\left[m/2t\cos\left(k_y/\sqrt{2}\right)\right]\left\{1+\left[
m/4t\cos\left(k_y/\sqrt{2}\right)\right]^2\right\}^{-1}$. If $m<4t
\cos(k_y/\sqrt{2})$, $\Theta$ satisfies the relation $\pi/2<\Theta
<\pi$, whereas for $m>4t\cos(k_y/\sqrt{2})$ one gets $0<\Theta<\pi
/2$. The narrow regions of $k_y$ near $k_y=\pm\pi/ \sqrt{2}$ turn
out not to be important for the effects in question and in the
estimations one can put $\cos(k_y/ \sqrt{2})\sim 1$.

The key properties of the odd S/AF/S junctions are associated with
the fact that they contain {\sl symmetric} magnetic interlayers
with respect to the two superconductors. In the (110) odd
interlayer both outermost chains belong to the majority spin
polarization, defined as ``spin-up'', and the Andreev states are
spin-split. The physical origin of the $\pi-0$ transition at
finite $T$ is just the interplay of spin-split Andreev bound
states. The general structure of the $\cal S$ matrix containing
the present reflection and transmission amplitudes, is similar to
that found in Ref. \onlinecite{bb02} for {\sl symmetric}
ferromagnetic interfaces. An important distinction to the AF case,
however, is the very different expression for the parameter
$\Theta$. The presence of the low-energy Q-reflection results in
quite large values $\Theta>\pi/ 2$, which ensure the $\pi-0$
transition with varying temperature in a wide range of the AF
order parameter $m<4t$. Applying the quasiclassical equations to
the superconducting leads and taking into account the AF interface
properties within the $\cal S$ matrix approach, we obtain the
following spectrum for the Andreev bound states:
$E^{\pm}_\sigma(k_y) =\sigma|\Delta|{\rm sgn}[\sin
\left(\frac{\Theta}{2}\pm\delta\right)]\cos\left(\frac{\Theta}{2}
\pm\delta\right)$. Here, $\sin\delta=\sqrt{D_{odd}}\cos(\chi/2)$
and $\chi$ is the phase difference between the left and right
superconducting leads. As seen from the energy spectrum
$E^{\pm}_\sigma(k_y)$, the parameter $m/4t$ becomes negligible in
tunnel junctions under quite strict conditions
$m/4t\ll\sqrt{D_{odd}}$. This implies $\Theta\approx\pi$ when four
dispersive bound states $E^{\pm}_\sigma(k_y)$ reduce to two doubly
degenerate states, leading to the particular results of
Ref.\onlinecite{bobkova05} for the Josephson current.

Below we consider the different conditions, $\sqrt{D_{odd}}\ll
m/4t,\,1$\, which can be easily satisfied in tunnel junctions. The
expression for the Josephson current which follows from the bound
state spectrum in the case $ \sqrt{D_{odd}}\ll m/4t,\, 1$,
describes the $\pi- 0$ transition with increasing temperature for
$\pi/2<\Theta<\pi$, and differs only by the additional factor of
$1/2$ from Eq. (4) of Ref. \onlinecite{bb02} (with $\alpha=1$)
valid for clean symmetric S/F/S tunnel junctions. At $T=0$ the
current is given by a sum over $k_y$ of the expressions
$j^{odd}_{c,k_y}(T=0)= -(1/2)e|\Delta|D_{odd}\cos(\Theta/2)$,
whereas near $T_c$ we get $j^{odd}_{c,k_y}(T\approx T_c)=-e
|\Delta|^2D_{odd}\cos\Theta/4T_c$. From this we see that the
product $j^{odd}_{c,k_y}(T=0)j^{odd}_{c,k_y}(T\approx T_c )$ is
negative for $\pi/2<\Theta<\pi$ and positive for $\Theta<\pi/2$.
Hence, if $\Theta<\pi/2$, there is no $\pi-0$ transition with
increasing $T$ and the $\pi$-state remains the ground state of the
odd (110) S/AF/S junctions for all $T<T_c$.

For the even (110) junctions, the outermost chains of the AF
interface have opposite spin polarizations and the Andreev states
are spin degenerate. In this geometry the spin-dependent
reflection amplitudes on the two sides of the interfaces differ
$r_{1,\sigma}=-r^*_{2,\sigma}=\sqrt{R_{ev}}\exp(i\sigma\Theta/2)$.
The transmission amplitude is real and spin-independent $d=\sqrt{
D_{ev}}=1/\cosh\left[2(L+1)\beta\right]$ and the expressions for
$\Theta$ and $\beta$ coincide with the odd case. Such structure of
the $\cal S$ matrix is characteristic also for the three-layer FIF
interface with antiparallel orientations of the two ferromagnetic
magnetizations \cite{bbk02}. The corresponding spin-degenerate
spectrum of the Andreev states in even S/AF/S junctions,
$E_{ev}(k_y)=\pm\Delta[D_{ev}(
k_y)\cos^2(\chi/2)+R_{ev}(k_y)\cos^2(\Theta/2)]^{1/2}$, has the
same form as that found for the FIF three-layer with antiparallel
magnetizations (see Eq.(4) in Ref. \onlinecite{bbk02}), but with a
significantly different expression for $\Theta$. The Josephson
current at zero temperature is a sum over $k_y$ of the expressions
$j^{ev}_{c,k_y}(T=0)=e|\Delta|D_{ev}/2\cos(\Theta/2)$, whereas
near $T_c$ we obtain $j^{ev}_{c,k_y}(T\approx T_c)=e|
\Delta|^2D_{ev}/4T_c$. There is no $0-\pi$ transition with varying
temperature, in agreement with our numerical results and with the
S/FIF/S junctions with antiparallel magnetizations.

The different symmetries of even versus odd AF interlayers are
also responsible for the low $T$ $0-\pi$ transitions as a function
of $L$. Indeed, the $\pi$-state is the ground state of clean S/F/S
junctions with $\alpha=1$, $\Theta>\pi/2$\cite{bb02}, whereas the
$0$-state is the ground state of the S/FIF/S junctions with
antiparallel magnetizations\cite{bbk02}. The direct analogy
between the odd (even) AF interfaces and the F interfaces at
$\alpha=1$ (the FIF interfaces with antiparallel magnetizations),
results in the correct sequence of the transitions at $T=0$. The
even/odd $0-\pi$ transitions can also be related, within the
perturbative approach, to effects of localized spin states in
interfaces and the anticommutation of
fermions\cite{shiba69,bulaev,glazman89,spivak91,choi00,mori05}.

The critical current $j_c$ in odd (110) S/AF/S junctions at $T=0$
is reduced by the factor $\cos(\Theta/2)$ compared to the standard
junctions with the transparency $D_{odd}$. For $m/t\ll 1$ we have
$\cos(\Theta/2)\ll 1$ and the relative suppression is significant.
The presumed condition $ \sqrt{D_{odd}} \ll m/4t$ gives the
smallest value $\cos( \Theta/2)\sim\sqrt{ D_{odd}}$ and a reduced
current $j^{odd}_{c} \propto D_{odd}^{3/ 2}$. Contrary to the odd
case, the critical current $j^{ev}_c$ in even junctions at $T=0$
is enhanced by $1/\cos(\Theta/2)$ compared to standard junctions
with the transparency $D_{ev}$. The maximal relative enhancement
is realized for $\cos(\Theta/2)\sim\sqrt{D_{ev}}$, where the
applicabilities of the present results border to those of Ref.
\onlinecite{bobkova05}. Then $j^{ev}_{c}\sim
e|\Delta|\sqrt{D_{ev}}$, in qualitative agreement with the net
critical current at $T=0$ found in Ref. \onlinecite{bobkova05}.

Fabrications of ultrathin interfaces with atomic-scale control of
the thickness over a macroscopic area, similar to the case of
ultrathin films\cite{guo04,bao05}, would allow observations of the
even/odd effect for S/AF/S junctions. Alternatively one should
average the Josephson current over interface imperfections. For
example, assume that the procedure can be reduced to an averaging
over thickness variations within a few layers. If $L\gg 1$, we get
$D_{odd}(L)\approx D_{ev} (L\pm 1)$. For $m\ll t$ the critical
currents are related as $j^{ev}_{c}\sim (t/m)^2| j^{odd}_{c}|\gg |
j^{odd}_{c}|$ at $T=0$, whereas near $T=T_c$ they are of the same
order and sign. Since the current $j^{ev}_{c}$ strongly dominates
$|j^{odd}_{c}|$ at low $T$, the result of the averaging is that
the ground state is the $0$-state with the anomalous critical
current $j^{ev}_{c}(T)$ and there is no $0-\pi$ transition, if
$m\ll t$. For $m\sim t$ even and odd currents are of the same
order and therefore similar samples can have differing signs of
$j_c$. If $\Theta<\pi/2$, $j^{ev}_{c}$ and $j^{odd}_{c}$ have
opposite signs and we expect a pronounced suppression of the net
critical current at all $T$. For $\Theta>\pi/2$, the odd and even
currents have opposite signs only at low $T$ where a similar
cancellation can take place. Therefore we predict a non-monotonic
temperature dependence of the net critical current in this case.

{\it Conclusions.} We have found a low temperature even/odd
sequence of the $0$- and $\pi$- states of S/AF/S junctions and an
additional novel $\pi-0$ transition with increasing temperature in
odd junctions. The even/odd effect is caused by the different
symmetries of the even versus odd AF interfaces and the
corresponding $\cal S$ matrices, and is revealed in qualitatively
different temperature dependencies of the critical currents for
even and odd barrier thicknesses. The $\pi-0$ transition with
varying temperature is induced in the odd junctions by the
interplay of the spin-split Andreev bound states.

{\it Acknowledgments.} This work was supported by ONR grant
N00014-04-0060 (B.M.A and P.J.H.), and by grants DOE
DE-FG02-05ER46236 (P.J.H and Yu.S.B.), NSF-INT-0340536 (I.V.B.,
P.J.H., and Yu.S.B.) and RFBR 05-02-17175 (I.V.B. and Yu.S.B).


\begin{thebibliography}{99}
%
\bibitem{sarma04}
I. Zuti\'c {\it et al}.,
Rev. Mod. Phys. {\bf 76}, 323 (2004).
%
\bibitem{efetov05}
F.~S.~Bergeret {\it et al}., Rev.~Mod.~Phys. {\bf 77}, (2005).
%
\bibitem{buzdin05}
A.~I.~Buzdin, Rev.~Mod.~Phys. {\bf 77}, 935 (2005).
%
\bibitem{ryazanov01}
V. V. Ryazanov {\it et al}., Phys. Rev. Lett. {\bf 86}, 2427
(2001).
%
\bibitem{aprili02}
T. Kontos {\it et al}., Phys. Rev. Lett. {\bf 89}, 137007 (2002).
%
\bibitem{beasley}
E. Terzioglu and M. R. Beasley, IEEE Trans. Appl. Supercond. {\bf
8}, 48 (1998).
%
\bibitem{blatter}
G.~Blatter {\it et al}.,
Phys. Rev. B {\bf 63}, 174511 (2001).
%
\bibitem{zagoskin}
A. M. Zagoskin, Physica C {\bf 368}, 305 (2002).
%
\bibitem{ustinov}
A.~V.~Ustinov, V.~K.~Kaplunenko,
J.~Appl.~Phys. {\bf 94}, 5405 (2003).
%
\bibitem{yamashita}
T. Yamashita {\it et al}., Phys. Rev. Lett. {\bf 95}, 097001
(2005).
%
\bibitem{bobkova05}
I. V. Bobkova {\it et al}., Phys. Rev.
Lett. {\bf 94}, 037005 (2005).
%
\bibitem{andersen05}
B. M. Andersen {\it et al}.,
Phys. Rev. B {\bf 72}, 184510 (2005).
%
\bibitem{BTK} G. E. Blonder, M. Tinkham, and T. M. Klapwijk, Phys.
Rev. B {\bf 25}, 4515 (1982).
%
\bibitem{bb02}
Yu. S. Barash and I. V. Bobkova, Phys. Rev. B {\bf 65}, 144502
(2002).
%
\bibitem{bbk02}
Yu. S. Barash {\it et al}.,  Phys. Rev. B {\bf 66},
140503 (2002).
%
\bibitem{shiba69}
H.~Shiba and T.~Soda, Progr.~Teor.~Phys. {\bf 41}, 25 (1969).
%
\bibitem{bulaev}
L. N. Bulaevskii, V. V. Kuzii, and A. A. Sobyanin, JETP Lett. {\bf
25}, 290 (1977).
%
\bibitem{glazman89}
L. I.~Glazman and K. A.~Matveev, JETP Lett. {\bf 49}, 659 (1989).
%
\bibitem{spivak91}
B.~I.~Spivak and S.~A.~Kivelson, Phys. Rev. B {\bf 43}, 3740 (1991).
%
\bibitem{choi00}
M.-S.~Choi {\it et al}.,  Phys. Rev. B {\bf 62}, 13569 (2000).
%
\bibitem{mori05}
M. Mori and S. Maekawa, Phys. Rev Lett. {\bf 94}, 137003 (2005).
%
\bibitem{guo04}
Y. Guo {\it et al}., Science {\bf 306}, 1915 (2004).
%
\bibitem{bao05}
X.-Y. Bao {\it et al}., cond-mat/0507172 (unpublished).
%
\end{thebibliography}
\end{document}